\documentclass[aps, prd, letterpaper, 
twocolumn, amsmath, amssymb, superscriptaddress,
floatfix, nofootinbib, longbibliography]{revtex4-1}
\usepackage[T1]{fontenc}
\usepackage{ragged2e}
\usepackage{float}
\usepackage[us,12hr]{datetime}
\usepackage{graphicx}
\usepackage{mathrsfs}
\usepackage{marvosym}
\newcommand{\comment}[1]{}
\usepackage{graphicx}
\usepackage{graphicx}
\usepackage{color}
\usepackage{bbold} 
\usepackage{xcolor}
\usepackage{mathrsfs}
\usepackage[braket, qm]{qcircuit}
\usepackage[colorlinks=true
,urlcolor=mn
,anchorcolor=black
,citecolor=magenta
,filecolor=navyblue
,linkcolor=rindou2
,menucolor=navyblue
,linktocpage=true
,pdfproducer=medialab]{hyperref}
\usepackage{slashed}

\usepackage{aas_macros,empheq,fancybox,graphicx,multirow}
\usepackage{subfig}
\usepackage[normalem]{ulem}

\definecolor{pc1}{rgb}{0.69, 0.25, 0.21}

\newcommand{\phase}[1]{\textcolor{blue}{{#1}}}

\definecolor{rindou1}{rgb}{0.4431,0.2862,0.7960}
\definecolor{rindou2}{rgb}{0.0078,0.1215,0.4392}
\definecolor{lapis}{rgb}{0.0.0470,0.2941,0.5568}
\definecolor{mn}{rgb}{0.15, 0.35, 0.95}
\newcommand\norm[1]{\left\lVert#1\right\rVert}

\usepackage{tikz,xcolor,hyperref}
\definecolor{lime}{HTML}{A6CE39}
\DeclareRobustCommand{\orcidicon}{%
	\begin{tikzpicture}
	\draw[lime, fill=lime] (0,0) 
	circle [radius=0.16] 
	node[white] {{\fontfamily{qag}\selectfont \tiny ID}};	\draw[white, fill=white] (-0.0625,0.095) 
	circle [radius=0.007];	\end{tikzpicture}
	\hspace{-2mm}}
\foreach \x in {A,...,Z}{%
	\expandafter\xdef\csname orcid\x\endcsname{\noexpand\href{https://orcid.org/\csname orcidauthor\x\endcsname}{\noexpand\orcidicon}}
}

\begin{document}
\title{Sachdev-Ye-Kitaev model on a noisy quantum computer}

\author{Muhammad Asaduzzaman~\orcidA{}\,}\email{muhammad-asaduzzaman@uiowa.edu}
\affiliation{Department of Physics and Astronomy, The University of Iowa, Iowa City, Iowa IA 52242, USA}
\author{Raghav G. Jha~\orcidB{}\,}\email{raghav.govind.jha@gmail.com}
\affiliation{Thomas Jefferson National Accelerator Facility, Newport News, VA 23606, USA}
\author{Bharath Sambasivam~\orcidC{}\,}\email{bsambasi@syr.edu}
\affiliation{Department of Physics, Syracuse University, Syracuse, NY 13244, USA}

\preprint{JLAB-THY-23-3980}

\begin{abstract}
We study the SYK model -- an important toy model for quantum gravity on IBM's superconducting qubit quantum computers. By using a graph-coloring algorithm to minimize the number of commuting clusters of terms in the qubitized Hamiltonian, we find the gate complexity of the time evolution using the first-order product formula for $N$ Majorana fermions is $\mathcal{O}(N^5 J^{2}t^2/\epsilon)$ where $J$ is the dimensionful coupling parameter, $t$ is the evolution time, and $\epsilon$ is the desired precision. With this improved resource requirement, we perform the time evolution for $N=6, 8$ with maximum two-qubit circuit depth of 343. We perform different error mitigation schemes on the noisy hardware results and find good agreement with the exact diagonalization results on classical computers and noiseless simulators. In particular, we compute vacuum return probability after time $t$ and out-of-time order correlators (OTOC) which is a standard observable of quantifying the chaotic nature of quantum systems. 
\end{abstract}

\maketitle
\section{Introduction}

\vspace{-3mm}

The holographic duality \cite{Witten:1998qj} relates a special class of quantum field theories in $d$ dimensions and quantum gravity in $d+1$ dimensions. This strong/weak duality enables one to study the properties of strongly coupled field theory using classical supergravity and vice versa. However, there are no cases where both sides of the duality can be studied analytically at the same time. Several attempts have been made on the lattice using Monte Carlo \cite{Catterall:2017lub,Catterall:2020nmn} to study these theories but they have their limitations. Therefore, it is often of interest to find simpler models that have holographic properties and can be studied in the strong coupling limit. One such model is the SYK model~\cite{Sachdev:1992fk,Kitaev:2015,Sachdev:2015efa,Maldacena:2016hyu} consisting of $N$ Majorana fermions in $0+1$ dimensions with random couplings between $q$ fermions at a time chosen from a Gaussian distribution with zero mean and variance proportional to $J^2/N^{q-1}$.

An interesting feature of the SYK model is that it develops an approximate conformal symmetry in the large $N$, low-temperature limit i.e., $N \gg \beta J \gg 1 $ ($\beta$ is the inverse temperature), where it is related to near extremal black holes that develop the nAdS$_{2}$ (near AdS$_2$) geometry. It was shown to saturate the chaos bound~\cite{Maldacena:2015waa}, a feature that is associated with holographic behavior. Since this is a 0+1-dimensional model, it is computationally tractable and has been studied up to 60 Majorana fermions ~\cite{Gur-Ari:2018okm, Kobrin:2020xms}. 

As a toy model for quantum gravity, it is, therefore, crucial to study the real-time dynamics of this model beyond methods accessible by classical computing. This direction has already been explored starting with Ref.~\cite{Garcia-Alvarez:2016wem}. In another work~\cite{Luo:2017bno}, the authors studied generalized SYK model using a four-qubit nuclear magnetic resonance (NMR) quantum simulator and computed bosonic correlation functions. 

We put forth an improved circuit complexity\footnote{The complexity is defined as the least number of two-qubit gates in the circuit that implements the time evolution of the Hamiltonian $H$} and study the SYK model on noisy superconducting quantum computers for the \emph{first} time. Specifically, we find an improved complexity from earlier proposals of $\mathcal{O}(N^{10}J^2 t^{2}/\epsilon)$ \cite{Garcia-Alvarez:2016wem} and $\mathcal{O}(N^{8}J^2 t^{2}/\epsilon)$ \cite{Xu:2020shn}
to $\mathcal{O}(N^{5} J^2 t^{2}/\epsilon)$ for the Lie-Trotter-based algorithm \cite{2019arXiv191208854C}. Using this improvement, we study the time evolution up to \emph{eight} Trotter steps on quantum hardware available through IBM and compute the return probability and four-point out-of-time-ordered correlators.

\section{\label{sec:SYK}SYK Hamiltonian}
The Hamiltonian for the SYK model with $N$ Majorana fermions and $q$-fermion interaction terms is: 
\begin{equation}
H = \frac{(i)^{q/2}}{q!} \sum_{i,j,k,\cdots ,q= 1}^{N} J_{ijk\cdots q}~~\chi_{i} \chi_{j} \chi_{k} \cdots \chi_{q},
\label{eq:SYK_main} 
\end{equation}
where $\chi$ are the Majorana fermions satisfying $\{\chi_i, \chi_j\} = \chi_i \chi_j + \chi_j \chi_i = \delta_{ij}$. We consider $q=4$ with random all-to-all quartic interactions averaged over disorder. The random (real) couplings $J_{ijkl}$ are sampled from a Gaussian distribution with the mean
$\overline{J_{ijkl}}=0$ and variance equal to 
$\overline{J_{ijkl}^{2}} = \frac{3! J^2}{N^3}$. We set $J=1$ in this work. The dimension of the Hilbert space is $\text{dim}(\mathcal{H}) = 2^{N/2}$ where $n = N/2$ is the number of qubits\footnote{We have $N = 2n$ since two Majorana fermions can be represented by 
one complex spinless fermion which can be represented by single qubit}. The model can also be considered for $q > 4$ and is solvable in the large $q$ limit \cite{Maldacena:2016hyu}.

\subsection{\label{subsec:QUB}Qubitization and Trotterization}

To perform the time evolution, we first have to map the fermionic Hamiltonian to qubits. We will use the standard Jordan-Wigner transformation. The $N$ fermions in \eqref{eq:SYK_main} can be written in terms of tensor product of $N$/2 Pauli matrices $X,Y,Z$ and the identity matrix $\mathbb{1}$ \cite{Garcia-Alvarez:2016wem,Luo:2017bno} as: 
\begin{align}
\chi_{2k-1} &= \frac{1}{\sqrt{2}} \Big(\prod_{j=1}^{k-1} Z_{j}\Big)X_{k} \mathbb{1}^{\otimes (N -2k)/2},  \nonumber \\
\chi_{2k} &= \frac{1}{\sqrt{2}} \Big(\prod_{j=1}^{k-1} Z_{j}\Big)Y_{k} \mathbb{1}^{\otimes (N -2k)/2},
\end{align}
where the square root is to ensure the normalization following Ref.~\cite{Kitaev:2015}. In order to simulate the dynamics on quantum hardware, we first decompose the Hamiltonian into Pauli strings as $H = \sum_{j=1}^{m} H_{j}$ and then use the standard Lie-Trotter 
formula \cite{Seth1996}: 
\begin{equation}
e^{-iHt} =  \Big(\prod_{j=1}^{m} e^{-iH_{j}t/r}\Big)^{r} + \mathcal{O}\Big(\sum_{j < k} \bigg\vert\bigg\vert [H_j, H_k] \bigg\vert\bigg\vert \frac{t^{2}}{r} \Big),
\label{eq:trotter_error}
\end{equation}
where we denote the spectral norm by $\norm{\cdot}$. If the terms in the decomposition of $H$ are near to commuting, then the Trotter error is reduced and vanishes if they commute. Though the number of terms into which $H$ is split is $m = {N \choose 4}$, one can reduce it by only summing over a small number of clusters of Pauli strings $\mathcal{N} \ll m$. This helps in reducing the Trotter error since the number of terms summed in \eqref{eq:trotter_error} is reduced. We find that for $N=6$, the reduction factor i.e., $m/\mathcal{N} = 3$ while for $N=8$, $m/\mathcal{N} = 35/3$ and this enables us to reliably evolve to larger times by controlling the Trotter error (see Supplemental Material for details). If we impose total error in simulating the time evolution is $\epsilon$, then we need $r = \mathcal{O}(t^2/\epsilon)$ Trotter steps assuming the spectral norm of commutators to be upper bounded by unity. Using these arguments, Ref.~\cite{Garcia-Alvarez:2016wem} estimated the circuit complexity of $\mathcal{O}(N^{10}t^{2}/\epsilon)$. The method based on Lie-Trotter expansion is not the only way of simulating Hamiltonians. Another way is to use a controlled version of oracles to embed the Hamiltonian in an invariant $SU(2)$ subspace \cite{Low:2016znh} and a variant of this was used in Ref.~\cite{Babbush:2018mlj} to bring complexity to $\mathcal{O}(N^{7/2}t + N^{5/2}t~\text{polylog}(N/\epsilon))$. However, this is not amenable to current hardware implementation. 

\begin{figure}
	\centering 
	\includegraphics[width=0.3\textwidth]{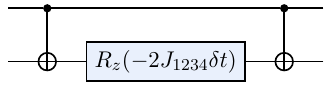}
	\caption{Circuit implementing single Trotter step for $N=4$.} 
 \label{fig:SYKcirc1}
\end{figure}

Let us consider the simplest case of $N=4$
where we have just one Pauli string with $H = -J_{1234} ZZ$. The time evolution circuit \cite{Kay:2018huf} is given by
Fig.~\ref{fig:SYKcirc1}.
\begin{table}
\setlength{\tabcolsep}{12pt}
\centering
\begin{tabular}{cccccc}
\hline
$N$ & Pauli strings & Clusters & Two-qubit gates \\
\hline
4 & 1 & 1 & 2 \\ 
\hline 
6 & 15 & 5 & 30 \\ 
\hline 
8 & 70 & 6 & 110 \\ 
\hline 
10 & 210 & 23 & 498 \\ 
\hline 
12 & 495 & 57 & 1504 \\ 
\hline 
14 & 1001 & 92 & 3560 \\ 
\hline 
16 & 1820 & 116 & 6812 \\ 
\hline
18 & 3060 & 175 & 11962  \\ 
\hline
20 & 4845 &  246 & 19984  \\  
\hline
\end{tabular}
\caption{\label{tab:tab1}The gate cost (assuming all-to-all connectivity) for the time evolution of the SYK Hamiltonian per Trotter step, number of Pauli strings, and the number of clusters $\mathcal{N}$ of commuting Pauli strings for different $N$.}
\end{table}
The two-qubit cost of simulating various $N$ can be found in Table~\ref{tab:tab1} for $N \le 20$. The circuit complexity estimate based on commuting Pauli strings exploiting the graph-coloring algorithm
grows\footnote{For $N \le 10$, the cost based on quantum Shannon decomposition (QSD) is lower~\cite{Shende2006}. However, large $N$ scaling is exponential.} like $\mathcal{O}(N^5)$, a substantial improvement over $\mathcal{O}(N^{10})$ proposed in Ref.~\cite{Garcia-Alvarez:2016wem}.  

The number of Pauli strings in the decomposition of $H$ 
grows like $\mathcal{O}(N^{4}/4!)$ for large $N$, and the cost to simulate the simplest non-trivial Pauli string is $\mathcal{O}(N)$. Hence, the $\sim N^5$ circuit complexity is close to \emph{optimal} for this approach to Hamiltonian simulation. The complexity can be improved by using Bravyi-Kitaev mapping, however, such optimizations are not required for the small $N$ which are currently hardware accessible. We provide details of gate resource estimation in the Supplemental Material (SM).

\section{Return probability}

One of the main motivations for using quantum computers is to understand the time evolution of quantum systems. An observable we compute is the return probability by evolving an initial state\footnote{The initial state chosen here belongs to the set of common eigenstates of the SYK spin operators defined in~\cite{Numasawa:2019gnl} and forms a complete basis} given by $ \ket{\psi_0} = \ket{0}^{\otimes n}$ for $N_{t}$ Trotter steps and computing the overlap as: 
\begin{equation}
    \mathcal{P}_{0} = \vert \langle \psi_{0} \vert e^{-iHt} \vert  \psi_{0} \rangle \vert^{2}. 
    \label{eq:ret_prob}
\end{equation}
The return probability is closely related to the spectral form factor \cite{Liu:2016rdi} and shows similar behavior of slope, dip, ramp, and plateau upon disorder average~\cite{Cotler:2016fpe,Numasawa:2019gnl} which are features of the SYK model. 

\begin{figure*} 
    \centering
    \includegraphics[width=1.\textwidth]{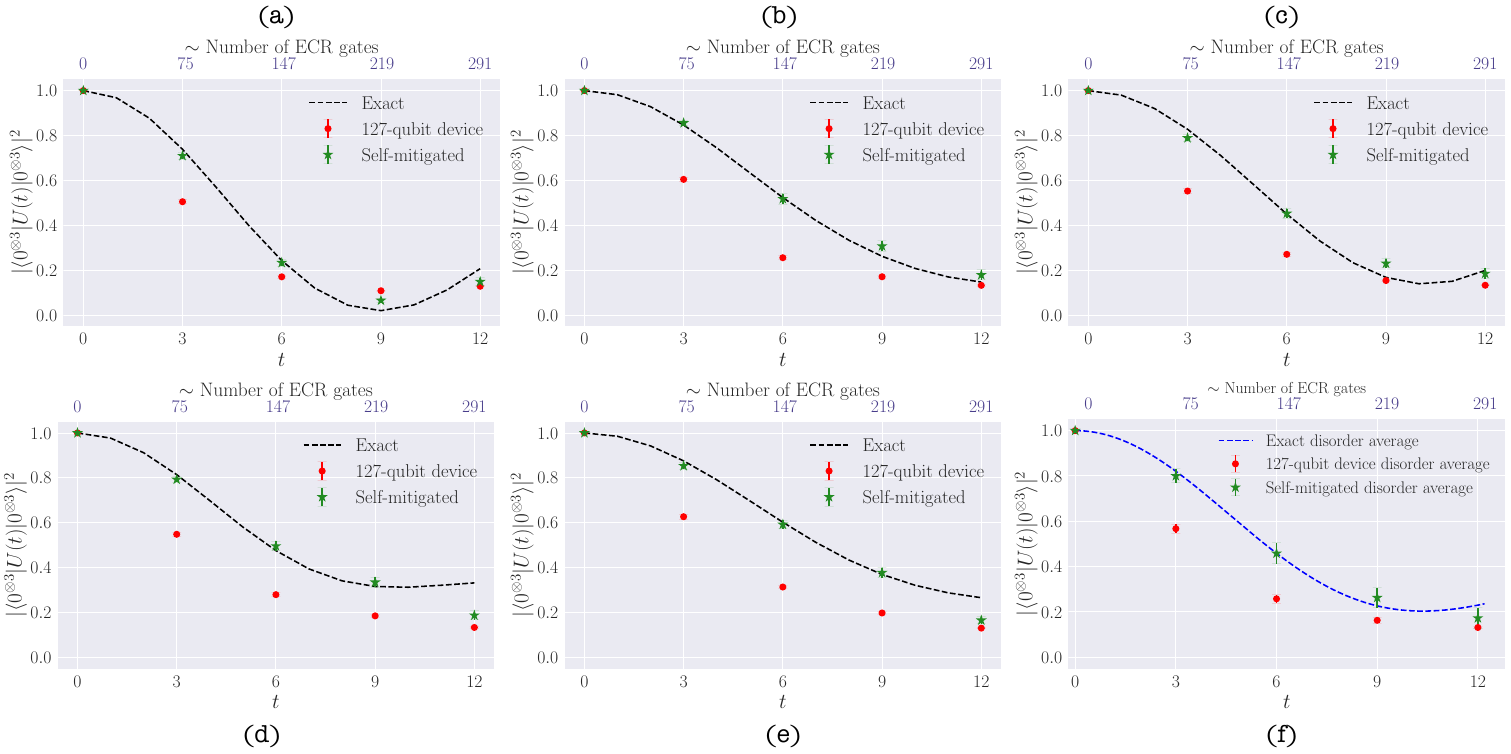}
    \caption{The return probability for five realisations (panels $a$ through $e$) of the SYK model with $N=6$. The disordered average (panel $f$) shows the slope region and the starting of the ramp behavior. The total two-qubit circuit depth is the same as the number of ECR gates, given the topology of the qubits.} 
    \label{fig:SYK_RP_hardwareN6}
\end{figure*}

The error-mitigated hardware results of the return probabilities for five realizations of the model with $N=6$ and disorder average are shown in  Fig.~\ref{fig:SYK_RP_hardwareN6}. The dashed black lines are the exact time evolution, while the dashed blue curve is the ensemble average of the exact evolution over the five realizations. At this stage, we do not compare to the state-of-the-art results obtained for the SYK model with classical computing methods since it is still far from the current best classical result with $N \sim 60$. 
\begin{figure}[htb] 
    \centering
    \includegraphics[width=0.4\textwidth]{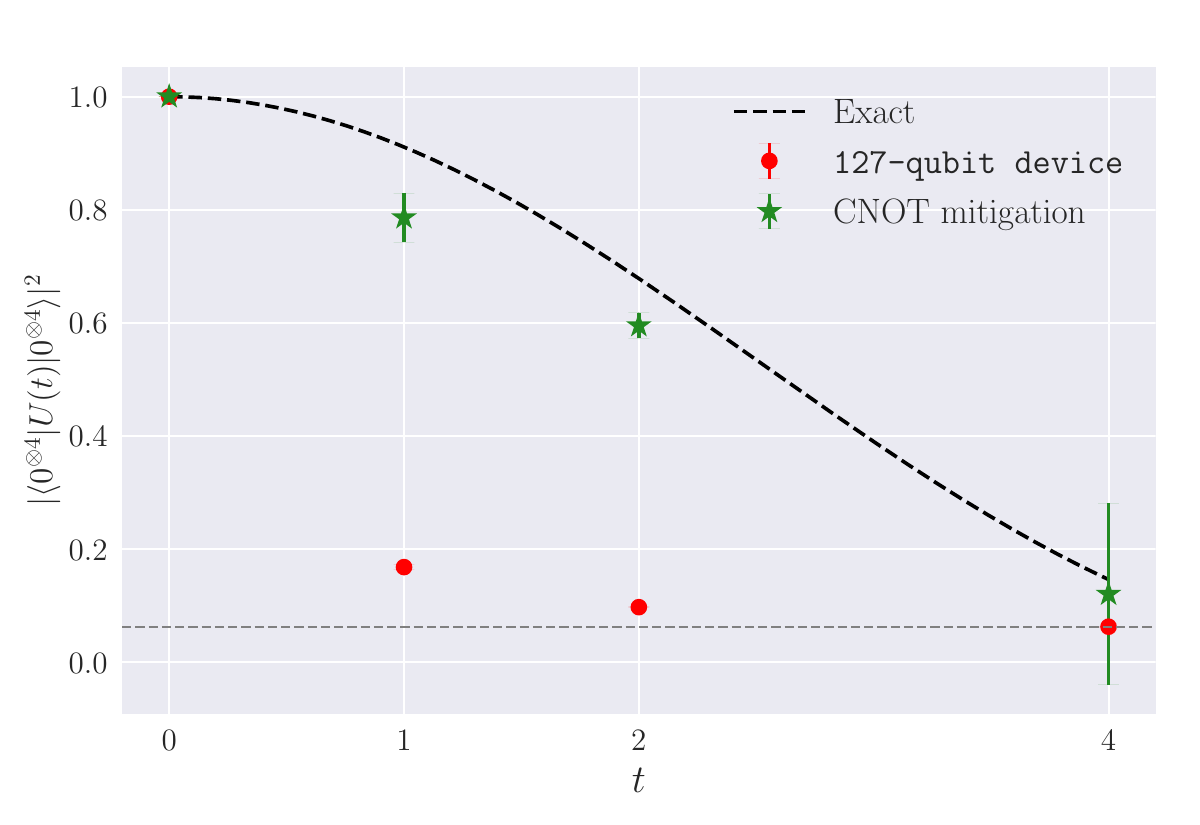}
    \caption{The return probability for a single instance of the SYK model with $N=8$. The results before mitigation \cite{Urbanek:2021oej} are slightly above the threshold of the fully depolarized channel (gray dashed line). For $t=1,2$ we use $dt=1$ while for $t=4$ we use
    $dt=2$. The maximum 2-qubit gate depth we use is $343$.}
    \label{fig:SYK_RP_hardwareN8}
\end{figure}

As $N$ increases, the resource requirements quickly increase as shown in Table~\ref{tab:tab1}. Therefore, for $N=8$ we consider the time-evolution of just one instance of the model in Fig.~\ref{fig:SYK_RP_hardwareN8}. The markers are results obtained from 127-qubit IBM machines \texttt{ibm\_cusco}, \texttt{ibm\_nazca}, and \texttt{ibm\_kyoto} with various degrees of error-mitigation applied. 

All these devices use the \texttt{eagle r3} processor, where the native two-qubit gate is not the standard CX but the echoed cross-resonance (ECR) gate (see SM for the definition of the gate). The leading source of gate noise in current devices is the two-qubit gates. To deal with this, the error mitigation strategy we employ is a combination of Pauli twirling/randomized compiling~\cite{Wallman_2016} for the ECR gates and self-mitigation~\cite{ARahman:2022tkr}. For the discussion on the implementation of these error-mitigation techniques and to get an estimate on the computation overhead to materialize in the quantum processing units of IBM, we refer the readers to the Supplemental Material (SM).
We also use the standard M3 protocol~\cite{PRXQuantum.2.040326} for qubit measurement errors and dynamical decoupling~\cite{PhysRevA.58.2733,ZANARDI199977,PhysRevA.59.4178,Ezzell:2022uat} to suppress decoherence noise. The results for the return probability\footnote{We only show even Trotter steps because for self-mitigation, one requires forward and backward Trotter evolution by an equal amount} with different types of error mitigation applied and exact evolution are shown in panels (\emph{a-e}) of Fig.~\ref{fig:SYK_RP_hardwareN6} for five different instances of the SYK Hamiltonian with $N=6$ and the disorder average over these realizations is shown in the last panel (\emph{f}). 

For $N=6$, the self-mitigated return probability agrees with exact results even for large evolution time and ECR gates in the circuit. To simulate the SYK model with $N=6$ up to $t=12$ (eight Trotter steps), we require about 300 ECR gates. For $N=8$, we need circuit depth of 343 ECR gates for two Trotter steps ($t=2$ with $dt=1$, $t=4$ with $dt=2$) and 170 (for $t=1$ with $dt=1$). The return probability for $N=8$ is shown in Fig.~\ref{fig:SYK_RP_hardwareN8} for one realization of the model. Going beyond $t=4$ appears to be past current hardware capability even with advanced error mitigation methods.

\section{OTOC computation} 

An important feature of the SYK model is that for large $N$ and in the low-temperature limit, it is maximally chaotic and is a quintessential example of a fast scrambler. A defining feature of such systems is that quantum information shared between a small number of elementary degrees of freedom is rapidly distributed into exponentially many degrees of freedom. This is known as `scrambling' and black holes are known to be the fastest scramblers in nature. To quantify the chaos, one considers the out-of-time (OTO) commutator between two operators $W$ and $V$ given by \cite{Hosur:2015ylk, Fu:2016yrv, Kitaev:2017awl, Lantagne-Hurtubise:2019svg}:
\begin{equation}
C(t) = -\langle [W(t), V(0)]^{\dagger}\,[W(t), V(0)] \rangle, 
\end{equation}
In general $C(t)$ starts from zero, and becomes significant at some later time $t$, which one refers to as scrambling time. By considering the commutator expansion, we can define the out-of-time order correlators (OTOC)\footnote{These were
first introduced in the study of disordered superconductors \cite{1969JETP...28.1200L}. OTOC is also closely related to the thermal average of signals from Loschmidt echo \cite{PhysRevLett.124.160603} 
}:
\begin{equation}
\label{eq:otoc1}
\text{OTOC} := F(t) = \langle W(t) V(0) W(t) V(0)\rangle_{\beta}, 
\end{equation}
where $W$ and $V$ are generic Hermitian operators and do not have the same symmetry as the Hamiltonian. The growth of the $C(t)$ is related to the decay of the OTOCs i.e., $F(t)$ through the simple relation, $C(t)  = 2 (1- F(t))$. The time-evolution of the operator $W$ in the Heisenberg representation is $W(t) = e^{iHt} W(0) e^{-iHt}$ and $\langle \cdot \rangle_{\beta} = \mathrm{Tr}\{\rho ~\cdot\}$ denotes the thermal average at inverse temperature $\beta$. A common choice for $\rho$ is to just use the $T \to \infty$ limit given by the normalized identity matrix, $\mathbb{1}/\text{dim}(\mathcal{H})$ \cite{Sunderhauf:2019djv}. For the $W$ and $V$ we can either take Pauli matrices or 
Majorana fermions such that $W = \chi_{i}$ and $V = \chi_{j}$ with $i \neq j$ and average over the different pairs $(i,j)$. We used the simplest case of a single Pauli matrix, i.e., $W = V = Z$. We denote $1 - F(t)$ by $1 - F(t)_{ij} = O_{ij}(t)$ where $i$ and $j$ denote the qubit location of the single-qubit operator $W$ and $V$ respectively and compute this on the hardware. The choice of these operators does not change the basic features of exponential growth and saturation. The SYK model saturates the chaos bound ~\cite{Maldacena:2015waa} at low temperatures and they have been extensively studied using classical computing methods \cite{Cao:2020rhe, Anegawa:2023vxq}. The current resources do not allow us to access very large values of $N$, but we take first step at computing in the simplest setting on the quantum computer.

\begin{figure*}
    \centering
    \includegraphics[width=0.8\linewidth]{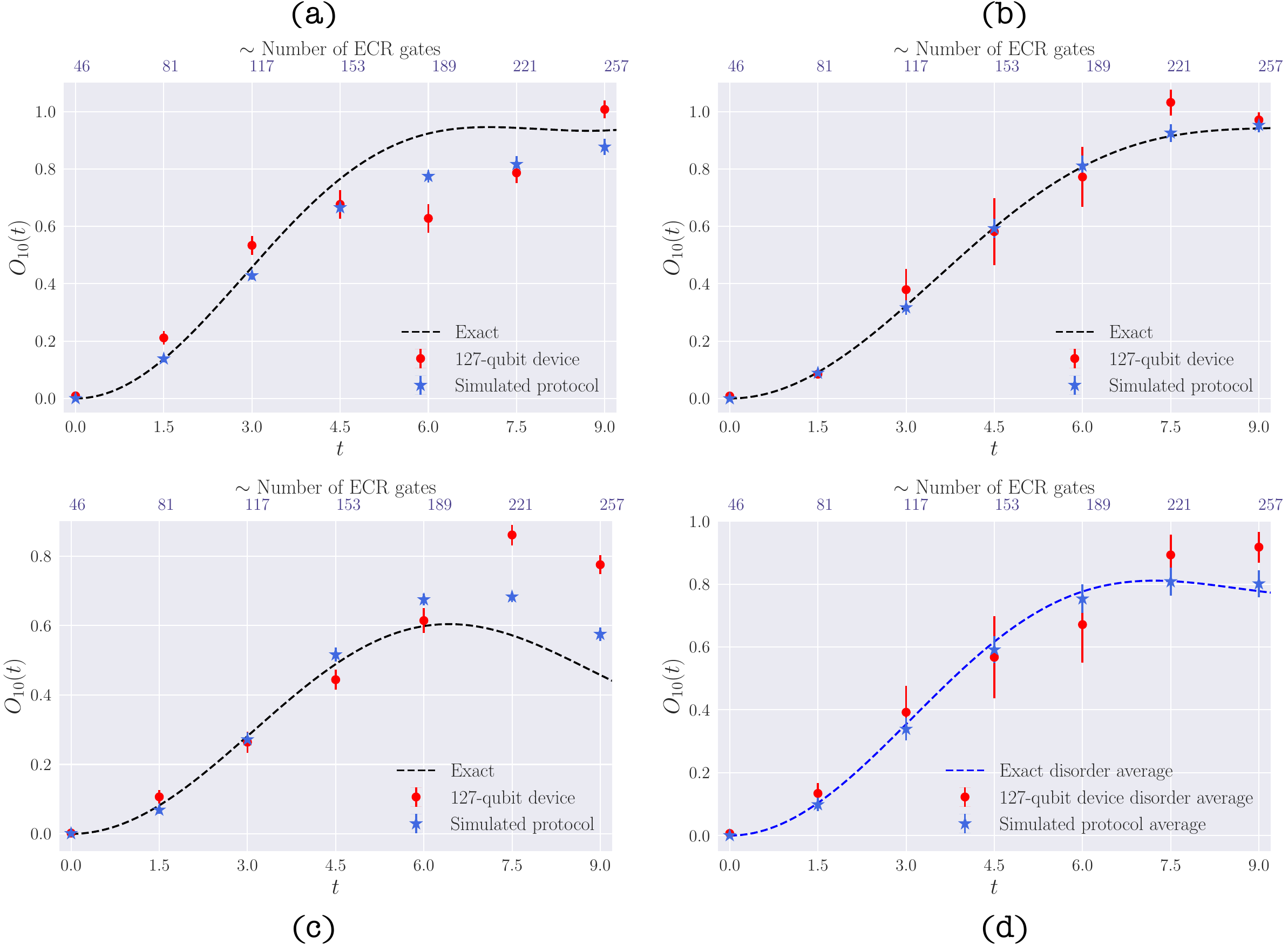}
    \caption{The OTOC for three realisations (panels $a$--$c$) of the SYK model with $N=6$ obtained on $\texttt{ibm\_kyoto}$ and $\texttt{ibm\_cusco}$.}
    \label{fig:otocN6}
\end{figure*}

\vspace{-3mm}
\subsection{Results for OTOC}
\vspace{-3mm}

Even though the OTOC seems simple to compute, the experimental/quantum computer measurement of the OTOC is very challenging because of the unusual time ordering. In a preliminary investigation, Ref.~\cite{PhysRevX.7.031011} studied OTOC of local operators on a nuclear magnetic resonance (NMR) quantum simulator followed soon on trapped-ion \cite{Garttner:2016mqj}. In order to compute OTOC on quantum hardware, we need to define a protocol for the measurement. Several proposals have been put forth \cite{Yao:2016ayk, 2018arXiv180600781A, Vermersch:2018sru, Mi:2021gdf} and we use the protocol proposed in Ref.~\cite{Vermersch:2018sru} which computes OTOC only by considering forward time evolution and exploiting the correlations between randomized measurements~\cite{Asaduzzaman:2023htk}. We discuss the details of the global protocol based on randomized measurements in the SM. 

The results for OTOC from \texttt{ibm\_cusco} and \texttt{ibm\_kyoto} for three instances of $N=6$ and the disorder average over them is shown in Fig.~\ref{fig:otocN6}. For this computation, we used the M3 readout error mitigation protocol and dynamical decoupling. Even without self-mitigation, we find good agreement with the exact results. 

\section{Summary and Discussion}

We have proposed circuit complexity of $\mathcal{O}(N^5 t^2/\epsilon)$ for Hamiltonian simulation of the SYK model with $N$ Majorana fermions, a substantial improvement over existing results and performed quantum simulations on noisy 127-qubit quantum computers. We studied the return probability for $N =6,8$ Majorana fermions and computed the out-of-time order correlators for $N=6$, a diagnostic of the chaotic behavior of quantum many-body systems. Due to the noisy devices currently available, we applied advanced mitigation methods to the hardware result and showed that it agrees well with the exact time evolution. 

It might appear that the superconducting platform is not the best method to study the quantum simulation of this model as we could have applied the ion-based approaches to quantum simulation. The advantage of superconducting platforms is the low gate times, but the limitation is the qubit connectivity. With ion-based platforms, this is the opposite- there is more freedom with connectivity, but the gate times are much longer. For the dense SYK model and other dense random Hamiltonians, both of these things are important. In this work, we take a step towards identifying which of these is more important by pushing the superconducting platform to push the limits with limited connectivity. We hope to extend this work with hardware admitting all-to-all connectivity.

Though we cannot study the strict holographic limit and see signs of saturation of chaos bound on current devices, we believe that our work will be useful in future explorations of this model. In this regard, it might also be useful to consider simplified models similar to SYK \cite{Xu:2020shn, Garcia-Garcia:2020cdo, Tezuka:2022mrr, Hanada:2023rkf} that are conjectured to have the same holographic behavior as pure SYK model considered here. Another direction is to consider $q > 4$ and explore the resource requirements and time evolution. It would be useful to study the dynamics of the model over different time scales for the return probability at finite $\beta$. These interesting problems would require resources that are beyond the contemporary hardware era. We leave these questions for future work. The use of quantum computers for models such as the SYK model in coming decades will not only provide new insights into the holographic principle but also into the interesting world of strange metals and quantum many-body systems \cite{Chowdhury:2021qpy}.

\subsection*{Acknowledgements}

MA is supported under the U.S. Department
of Energy grant DE-SC0019139. RGJ is supported by the U.S. Department of Energy, Office of Science, National Quantum Information Science Research Centers, Co-design Center for Quantum Advantage (C2QA) under contract number DE-SC0012704 and by the U.S. Department of Energy, Office of Science, Office of Nuclear Physics under contract number DE-AC05-06OR23177. BS is supported in part by the U.S. Department of Energy, Office of Science, Office of High Energy Physics, under Award Number DE-SC0009998. MA and BS would like to thank Jefferson Lab's Quantum Computing Bootcamp for the hospitality where this work started. We thank the IBM quantum hub at Brookhaven National Laboratory for providing access to the IBMQ quantum computers on which the computations were done. 

\subsection*{Data Availability Statement}
The $N=6, 8$ SYK Hamiltonian realizations, the Pauli decomposition, and the time evolution circuit as $\textsc{open qasm 2.0}$ files for single Trotter step can be obtained from Ref.~\cite{Asad_SYK_2023}. 
\bibliographystyle{utphys}

\appendix
\onecolumngrid


\section{Optimizing the time evolution circuit and return probability}
\justify 
One of the main results of the paper is how to perform efficient time evolution of the model. We now provide details related to optimizing the gate costs of the time evolution based on commuting Pauli strings in the decomposition of the qubit Hamiltonian. To obtain this improved scaling, we used well-known methods such as partitioning of the Pauli terms in commuting clusters (using the degree of saturation at each node i.e., DSatur \cite{10.1145/359094.359101} graph-coloring  algorithm) and then constructing the diagonalizing operator (circuit) for each cluster \cite{van_den_Berg_2020, 2020arXiv200105983G, Miller:2022sol, kurita2022pauli, Murairi:2023oti} separately followed by exponentiation for time evolution. At the end we add the gate costs from all clusters \cite{Jena2019, Verteletskyi_2020, Murairi:2022zdg}.

\begin{figure}[h!]
	\centering 
	\includegraphics[width=0.8\textwidth]{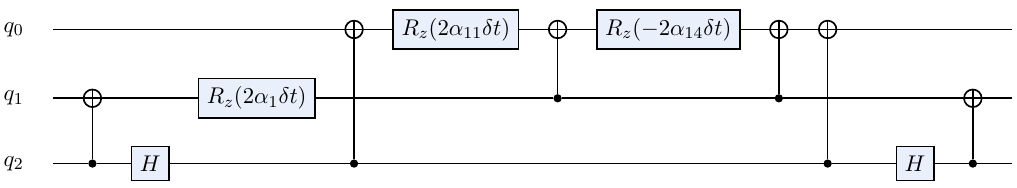}
	\caption{The circuit to implement $\exp(-iH_{1}\delta t)$ for $H_1$ given in \eqref{eq:H1cluster} for one Trotter step. The CX and H gates on the left and right end are part of the diagonalizing unitary circuit. The total cost is 6 CX gates as mentioned in the text.}
 \label{fig:example1}
\end{figure}

Let us take $N=6$ as an example. In this case, we have a three-qubit Hamiltonian which can be decomposed into 15 Pauli strings each with random coupling taken from a well-defined distribution. The generic Hamiltonian is written as: 
\begin{align}
H = \alpha_{1}\mathbb{1}ZZ + \alpha_{2}X\mathbb{1}X + \alpha_{3}X\mathbb{1}Y + 
 \alpha_{4}XXZ +  \alpha_{5}XYZ + 
  \alpha_{6}Y\mathbb{1}X +
   \alpha_{7}Y\mathbb{1}Y +
    \alpha_{8}YXZ +  \alpha_{9}YYZ
    +  \alpha_{10}Z\mathbb{1}Z \\ \nonumber 
    +  \alpha_{11}ZXX
    +  \alpha_{12}ZXY \newline
    +  \alpha_{13}ZYX 
    +  \alpha_{14}ZYY
    + \alpha_{15}ZZ\mathbb{1},
\end{align}
where $\alpha_{1, 2, \dots, 15}$ are the random couplings for a single realization of the model. By inspection, we find that it is possible to divide the terms of $H$ into five clusters where all Pauli strings in a given cluster commute with each other. The clusters are:
\begin{equation}
  \{\mathbb{1}ZZ,ZXX,ZYY\},\{X\mathbb{1}X,YXZ,ZXY\},\{X\mathbb{1}Y,Y\mathbb{1}X,Z\mathbb{1}Z\}, \{XXZ,YYZ,ZZ\mathbb{1}\},\{XYZ,ZYX,Y\mathbb{1}Y\}. 
\end{equation} 
Since they commute, they can be simultaneously diagonalized. The diagonalizing unitary circuit for each cluster $V$, satisfies $V^{\dagger} H_{p} V = \widetilde{H}_{p}$ where $p=1 \cdots 5$ and $\widetilde{H}_{p}$ is diagonal and can be written completely in terms of $Z$ and $\mathbb{1}$.
Therefore, rather than doing $\exp(- i H_{p} \delta t)$, one performs $\exp(-i (V \widetilde{H}_{p} V^{\dagger}) \delta t)$. By making use of the well-known identity: 
\begin{equation}
    Z \otimes Z = \text{CNOT} \cdot (\mathbb{1} \otimes Z) \cdot \text{CNOT},
\end{equation}
we can change two $Z$'s to a single $Z$ and by repeatedly applying this, we can reduce any Pauli string $\widetilde{H}_{p}$ to single $Z$. Once we have single $Z$, then we can use the identity that $e^{it I \otimes \mathbb{P} \otimes I} = I \otimes e^{it \mathbb{P}} \otimes I$, where $\mathbb{P}$
is a single Pauli matrix (X, Y, or Z) because $
e^{it I \otimes \mathbb{P} \otimes I} = \cos(t) I \otimes I \otimes I + i\sin(t) I \otimes \mathbb{P} \otimes I = 
 I \otimes \big( \cos(t) I + i \sin(t) \mathbb{P}  \big) \otimes I = I \otimes e^{i \mathbb{P} t} \otimes I$. Therefore, 
 $\exp(-it I \cdots I Z \cdots I \cdots I) = I \cdots I R_{z}(2t) \cdots I \cdots I$. As an example, for the first cluster i.e., 
 \begin{equation}
 \label{eq:H1cluster} 
  H_{1} = \alpha_{1} IZZ + \alpha_{11} ZXX + \alpha_{14} ZYY,   
 \end{equation}
we have the diagonalizing circuit made up of one CX and one Hadamard gate. It is easy to show that: 
\begin{equation}
  \widetilde{H}_{1} = \alpha_{1} IZI + \alpha_{11} ZIZ - \alpha_{14} ZZZ.  
\end{equation}
For the first term, we do not need any CX gate, just the rotation $R_{z}$ gate. For the second term, we need two CX for the first and third qubit. For the last term, one would naively expect four CX gates from the basic rule that one needs $2(\nu - 1)$ CX for the Pauli term with Hamming weight $\nu$, but since the terms commute, we can use two CX gates for the second term to bring down the required total cost to 4 CX gates. We need two CX gates for $V$ and $V^{\dagger}$ adding to a total of 6 CX gates. The circuit for one Trotter step is shown in
Fig.~\ref{fig:example1}.
The next four clusters require 6, 4, 6, and 8 gates respectively, bringing the total cost for $N=6$ to 30 CX gates as mentioned in the main text for a single Trotter step. If we take the first cluster and do not use the information that can be simultaneously diagonalized, it will increase the circuit complexity. For example, if we just use the optimization levels available in $\textsc{QISKIT}$, the cost of time evolution for \eqref{eq:H1cluster} will be 17 CX gates while is about three times more than our estimate. We do a similar procedure for $N=8$ and find that 70 Pauli terms can be organized into six commuting\footnote{Checking whether two Pauli strings of length $n$ commute is straightforward. One needs to check the commutativity of two strings qubit-wise. If the number of qubits denoted by $k$ where they don't commute is even, the strings commute. Let us consider two six-qubit ($n=6$) Pauli strings, $XYZZXY$ and 
$YX\mathbb{1}ZXY$. Only the qubits at the first ($X$ and $Y$) and second ($Y$ and $X$) locations don't commute i.e., $k = 2$. Therefore, the strings commute. It is easy to check that $XYZZXY$ and $YX\mathbb{1}ZXZ$ don't commute since $k = 3.$} clusters each with 14, 12, 8, 14, 12, and 10 terms as given below:
\begin{align}
\label{eq:N8PS}
\{\mathbb{1}\mathbb{1}ZZ,ZZ\mathbb{1}\mathbb{1},\mathbb{1}ZXX,Z\mathbb{1}YY, \mathbb{1}ZYY, Z\mathbb{1}XX, XX\mathbb{1}Z,YYZ\mathbb{1},XXZ\mathbb{1},YY\mathbb{1}Z,
XYYX,YXXY,XYXY,YXYX
 \} \nonumber \\
\{\mathbb{1}XYZ,ZYX\mathbb{1},\mathbb{1}Y\mathbb{1}Y,ZXZX, \mathbb{1}ZYX, Z\mathbb{1}XY, X\mathbb{1}ZY,YZ\mathbb{1}X ,XZXZ,Y\mathbb{1}Y\mathbb{1}, XYZ\mathbb{1},YX\mathbb{1}Z \}
\nonumber \\ 
 \{XYYY, YXXX, XYXX, YXYY, XXXY, YYYX, XXYX, YYXY\}
 \nonumber \\ 
    \{\mathbb{1}X\mathbb{1}X,ZYZY, \mathbb{1}YXZ, ZXY\mathbb{1}, \mathbb{1}ZXY, Z\mathbb{1}YX, X\mathbb{1}X\mathbb{1}, XZ\mathbb{1}Y, Y\mathbb{1}ZX, YZYZ, XY\mathbb{1}Z, YXZ\mathbb{1}, XXXX, YYYY\} \nonumber \\
    \{\mathbb{1}X\mathbb{1}Y, ZYZX, \mathbb{1}Y\mathbb{1}X, ZXZY, \mathbb{1}Z\mathbb{1}Z, Z\mathbb{1}Z\mathbb{1}, X\mathbb{1}Y\mathbb{1}, XZYZ, Y\mathbb{1}X\mathbb{1}, YZXZ, XXYY, YYXX\} \nonumber \\
   \{\mathbb{1}XXZ, ZYY\mathbb{1}, \mathbb{1}YYZ, ZXX\mathbb{1}, \mathbb{1}ZZ\mathbb{1}, Z\mathbb{1}\mathbb{1}Z, X\mathbb{1}ZX, XZ\mathbb{1}X, Y\mathbb{1}ZY, YZ\mathbb{1}Y\} \nonumber. \\
\end{align}

\begin{figure} 
 \vspace{4mm}
	\centering 
	\includegraphics[width=0.6\textwidth]{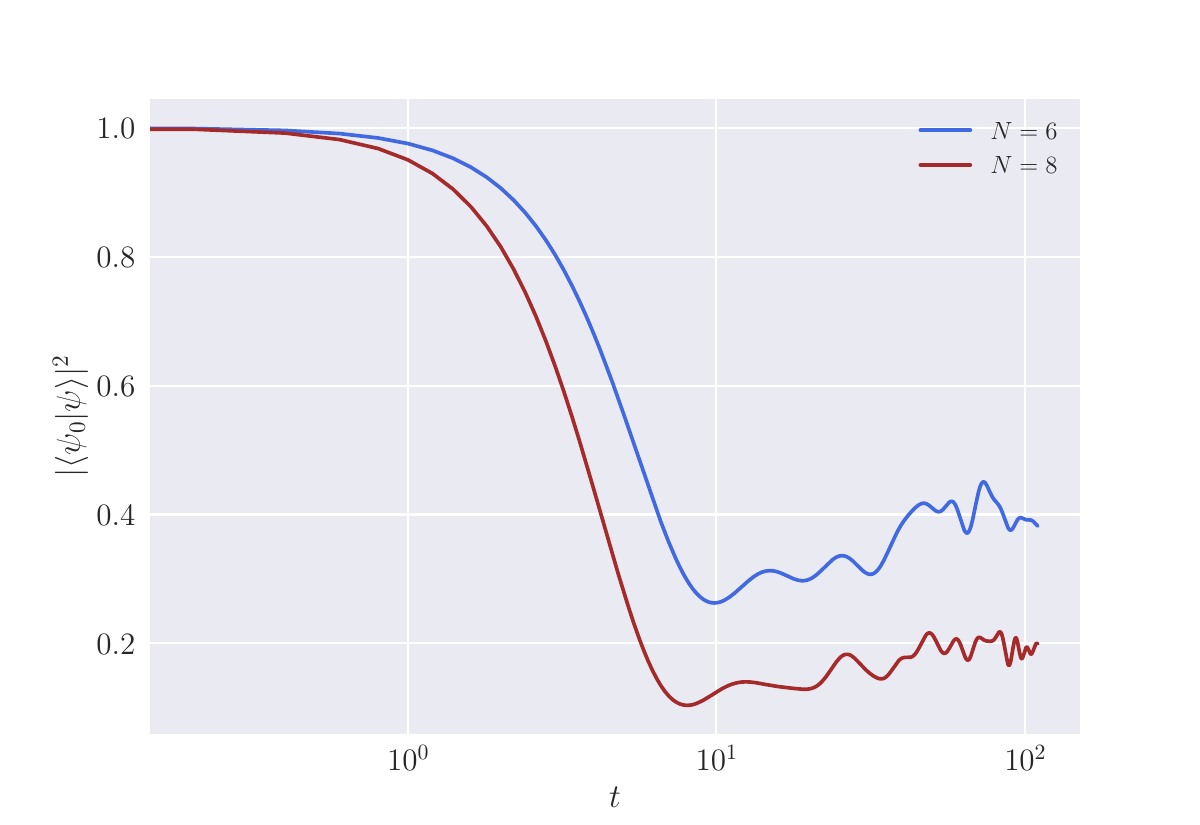}
	\caption{\label{fig:SYK_rp100inst}The exact result for return probability as a function of evolution time $t$ with 100 instances of the SYK model. The initial state is taken to be $\ket{0}^{\otimes n}$. The plateau value $\sim 1/L$ where $L = \text{dim}(\mathcal{H}) = 2^n$ is the dimension of the Hilbert space.} 
\end{figure}

The quantum circuit for the time evolution of $N=8$ SYK Hamiltonian needs 22, 20, 14, 18, 16, and 20 CX gates per cluster of \eqref{eq:N8PS} summing to a total of 110 CX gates per Trotter step. After optimizing the time evolution circuits with $\delta t$, we now need to make a good choice of the time step. For this, we did extensive tests of the observables on noise model devices and noiseless simulators and found that $\delta t = 1.5$ is a reasonably good choice. Our results for both return probability and OTOC are in good agreement with this step size. One of the observables we compute on quantum hardware is the return probability 
Since we focus on the infinite temperature limit ($\beta = 0$) in this work, we do not have access to the standard spectral form factor (SFF). However, for the SYK model, certain features of the SFF are also seen in the return probability as explored in Ref.~\cite{Numasawa:2019gnl}. We compute the return probability for different $N$ using exact time evolution for $N=6,8$. Similar to SFF, one expects to see various stages (referred to as slope and ramp) before it converges to a constant value referred to as a `plateau'. The plateau value is the fraction of the initial state retained after sufficiently long time evolution and this depends on the size of the Hilbert space. The slope region is self-averaging (i.e., a single instance of the model shows the same qualitative behavior as the disordered averaged case) while the ramp and plateau regions for SFF and return probability are not self-averaging. We show the exact results for 100 instances of the evolution in Fig.~\ref{fig:SYK_rp100inst}. The slope region for $N =6$ ends around $t=10$ and this is consistent with the disorder average over \emph{only} five instances from the hardware results 
in the main text. 

\section{\label{app:twirling}Pauli twirling}

\renewcommand{\arraystretch}{1.5}
\begin{table}[h!]
\centering
\begin{tabular}{||c||c|c|c|c|c|c|c|c|c|c|c|c|c|c|c|c||}
     \hline
     $P_1$ & $\mathbb{1}$ & X & \phase{X} & $\mathbb{\phase{1}}$ & Y & Z & \phase{Z} & \phase{Y} & Z & \phase{Y} & $\phase{\mathbb{1}}$ & $\mathbb{\phase{1}}$& Y & \phase{Z} & \phase{X} & \phase{X}  \\
     \hline
     $P_2$ & $\mathbb{1}$ & $\mathbb{1}$ & \phase{Z} & \phase{Z} & X & X & \phase{Y} & \phase{Y} & $\mathbb{1}$ & $\mathbb{\phase{1}}$ & \phase{X} & \phase{Y}& Z & \phase{Z} & \phase{Y} & \phase{X}  \\
     \hline
     $P_3$ & $\mathbb{1}$ & X & \phase{X} & $\mathbb{\phase{1}}$ & Y & Z & \phase{Z} & \phase{Y} & Y & \phase{Z} & \phase{X} & \phase{X}& Z & \phase{Y} & $\mathbb{\phase{1}}$ & $\mathbb{\phase{1}}$  \\
     \hline
     $P_4$ & $\mathbb{1}$ & $\mathbb{1}$ & \phase{Z} & \phase{Z} & X & X & \phase{Y} & \phase{Y} & Z & \phase{Z} & \phase{Y} & \phase{X}& $\mathbb{1}$ & $\mathbb{\phase{1}}$ & \phase{X} & \phase{Y}  \\
     \hline
\end{tabular}
\caption{The 16 conjugations that leave the ECR gate invariant (see Fig.~\ref{fig:pauli_twirl}). The conjugations in blue yield an overall phase of $\pi$.}
\label{table:ECR_conj}
\end{table}

The usefulness of the noisy intermediate-scale quantum (NISQ) devices is limited by the noise and error that are associated with them. Some common types of errors are — readout errors and gate errors. Readout errors occur when we measure the state of the qubit incorrectly resulting in different probabilities and expectation values. The gate errors are due to the application of gates in a quantum circuit. They are usually classified into coherent and incoherent errors. One useful way of differentiating between coherent and incoherent error is how the infidelity increases with circuit size or the number of times the channel is applied. If it increases linearly, it is a type of incoherent noise. The coherent errors add up quadratically \cite{2020NJPh...22g3066I}. The coherent errors preserve state purity such that a state that is lying on the surface of the Bloch sphere does not leak in the radial direction and stays pure (on the surface). Incoherent errors map pure states to mixed states such that they can be represented by states inside the Bloch sphere. The incoherent errors are due to the interaction with the outside environment. The incoherent (stochastic) errors are often modeled by the depolarising noise model given by \cite{Urbanek:2021oej}:
\begin{equation}
\mathcal{E}(\rho) = (1-p) \rho + p \mathbb{1}/2^{n}, 
\end{equation}
where $\mathcal{E}$ denotes the completely positive trace-preserving map (CPTP) noise channel and $\rho$ is the density matrix and $p$ is the error rate also known as depolarising parameter. If we consider a $n$-qubit channel, then a probability of $p = 1 - (1/2^n)$ is referred to as `completely depolarizing' or `pure noise' since it destroys the polarization of the qubit and any pure state is changed to a maximally mixed state. The expectation value of an operator $\mathcal{O}$ for the depolarizing channel is given by:
\begin{equation}
\overline{\langle O \rangle} = \text{Tr}(\mathcal{E}(\rho) O) = (1-p) \langle O \rangle + \frac{p}{2^n} \text{Tr}(O)
\end{equation}
The dominant gate errors are coherent (do not spoil unitarity) in nature and hence are not explained by the 
depolarizing noise channel. However, one can use randomized compiling to transform them into incoherent errors. 
We do this by inserting a pair of single-qubit gates on both sides such that the two-qubit gate remains unchanged. 
These possible choices for the ECR gate are shown in Table~\ref{table:ECR_conj}. The basis idea of Pauli twirling (also sometimes known as `randomized compiling') \cite{He:2020udd, Urbanek:2021oej} is to do operations on two-qubit gates by inserting one-qubit gates from Pauli groups on either side (conjugation) such that the initial gate is left unchanged. The only two-qubit gate we have used is the echoed cross-resonance (ECR) gate
which implements: 
\begin{equation}
    \text{ECR}~q_0, q_1 = \frac{1}{\sqrt{2}}(IX - XY)  =
\frac{1}{\sqrt{2}}
\begin{pmatrix}
    0 & 0 & 1 & -i \\
    0 & 0 & -i & 1 \\
    1 & i & 0 & 0 \\
    i & 1 & 0 & 0
\end{pmatrix}.
\label{eq:def-ECR}
\end{equation}
\begin{figure}[H]
    \centering
    \includegraphics[width=0.6\linewidth]{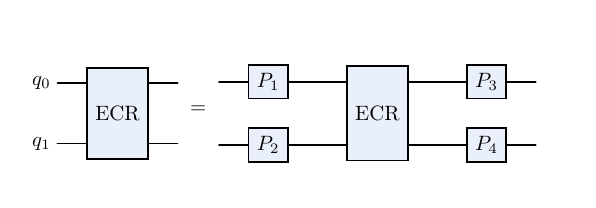}
    \caption{The circuit implementing the twirling of the ECR gate by choosing the $P$ gates from the Pauli group. This is also sometimes referred to as `randomized compiling'. This procedure changes the realistic noise which is coherent in nature to stochastic noise.}
    \label{fig:pauli_twirl}
\end{figure}

 For the ECR gate, there is a set of 16 single-qubit conjugations that leave the ECR gate invariant, up to a global phase. For each gate implementation, the noise is `twirled' differently for different conjugations. This helps convert a majority of the coherent noise in the circuit into stochastic noise that is well-modeled by the depolarizing quantum channel. We show this in Fig.~\ref{fig:pauli_twirl}. One of them is the trivial conjugation ($P_{1,2,3,4} = \mathbb{1}$). All conjugations ($P_{1} \otimes P_{2} \cdot \text{ECR} \cdot P_{3} \otimes P_{4}$) are listed below:

\section{\label{app:self-mitigation}Self-Mitigation}
We construct mitigation circuits with the same structure and gate count as the physics circuits, to characterize the probability of errors $p$ due to the noise. For self-mitigation, we construct a circuit consisting of $N_t/2$ Trotter steps forward in time and $N_t/2$ Trotter steps backward in time. On a noiseless (purely unitary) device, this would bring us back to the initial $\ket{\psi_0}$ state, up to Trotter error. However, on a noisy device, the state after the mitigation circuit would be some $\ket{\overline{\psi_0}}$. The probability of error is then $p= 1-\vert \bra{\psi_0}\overline{{\psi_0}}\rangle \vert^2$. Using this, we can extract the noiseless return probability $\langle \mathcal{P}_{0}\rangle$ from the noisy one $\overline{\langle \mathcal{P}_{0}\rangle}$ using~\cite{PhysRevLett.127.270502}
\begin{equation}
    \langle \mathcal{P}_{0}\rangle = \frac{\overline{\langle \mathcal{P}_{0}\rangle} - {2^{-n}}p}{1-p}.
    \label{eq:noiseless_exp}
\end{equation}
We build 75 Pauli-twirled physics and self-mitigation circuits each and execute both of them for 2048 shots.

\section{Protocol for OTOC computation}



For the computation of OTOC 
, we have to make a choice of operators, $W$ and $V$. However, the choice of operators in the four-point function does not change the generic behavior of the growth of OTOC \cite{PhysRevLett.124.160603}. We studied the exact time evolution using different choices of operators and found that the qualitative difference is not substantial. Due to the limited quantum resources in implementing the OTOC protocol and for benchmarking our results, we have restricted to the simplest choice of $W$ and $V$ i.e., just one non-trivial Pauli matrix in the three-qubit Pauli string.

We follow the Ref.~\cite{Vermersch:2018sru} where they discussed a `global protocol' to compute OTOC which uses randomly generated unitaries from the circular unitary ensemble (CUE). By applying random unitaries, many random states are created to mimic a thermalized scenario for the computation of the OTOC. The protocol is developed using the following equation for traceless operators~\cite{chen2018measuring}, 
\begin{equation}
    \operatorname{Tr}\left[W(t) V^{\dagger} W(t) V\right]  = \frac{1}{ N_{\mathcal{H}}(N_{\mathcal{H}}+1) } \overline{\langle W(t)\rangle_{u, k_0}\left\langle V^{\dagger} W(t) V\right\rangle_{u, k_0}} . \label{eqn_trace_ensemble}
\end{equation}

\begin{figure}[H]
    \centering
    \includegraphics[width=0.9\linewidth]{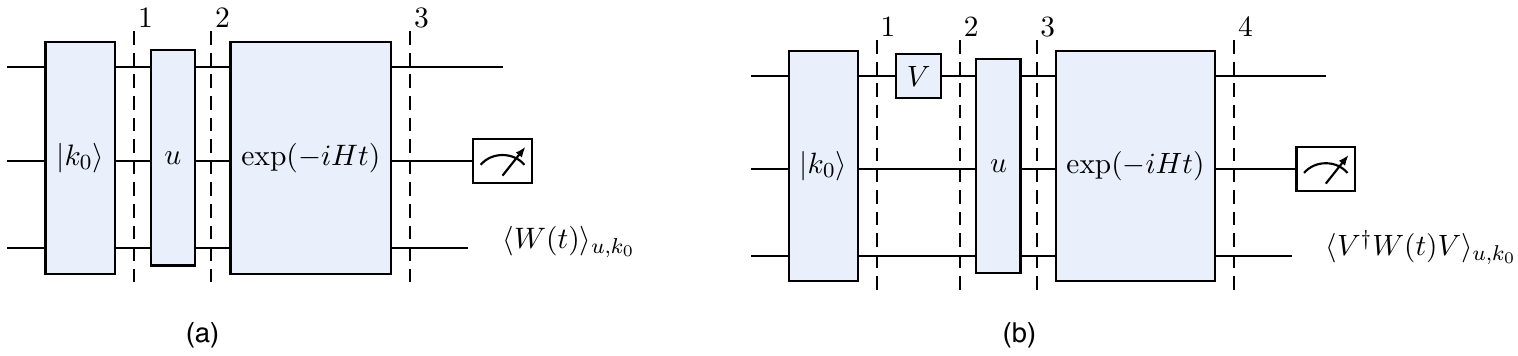}
    \caption{OTOC is computed from the correlation of the measurement of two different operators (a) $\langle W(t) \rangle $ and (b) $\langle V^\dagger W(t) V \rangle$. The same set of unitaries $\{u_{1},\cdots\ u_{N_u}\}$ are required to find the correlation between the expectation value of the operators.}
    \label{fig_OTOC_global_protocol}
\end{figure}

On the right-hand side, the overline denotes an ensemble average of measurements over a set $u=\{u_{0},u_{1},\cdots u_{N_{u}}\}$ of random unitary operators, $k_0$ denotes an arbitrary initial state and $N_{\mathcal{H}}$ is the dimension of the Hilbert space. The implementation of 
the global protocol requires creating a $n$-qubit random unitary operator that applies to the input state $\ket{k_0}$. 
The decomposition of an $n$-qubit random unitary in terms of two-qubit gates scales exponentially with $n$ \cite{Shende2006}. For a small number of qubits, the gate cost is manageable with the current NISQ-era machine. We outline the steps to compute OTOC by the randomized protocol developed in Ref.~\cite{Vermersch:2018sru} below:

\begin{itemize}
	\item  We prepare an arbitrary initial state $|k_0\rangle$ (position 1 in Fig.~\ref{fig_OTOC_global_protocol}(a)). For simplicity, we pick the state $\ket{\mathbf{0}}=\ket{0}^{\otimes n}$ similar to the calculation of the return probability. Next, we apply a $n$-qubit random unitary sampled uniformly from the CUE which results in a random state $|\psi_1\rangle = u \ket{\mathbf{0}} $ (position 2 in Fig.~\ref{fig_OTOC_global_protocol}(a)).  
	\item  The Trotter evolution of the random state is computed using the Trotterized time-evolution operator $U(N_t)=\Big[\exp(-i \hat{H} \delta t) \Big]^{N_{t}}$. This yields $|\psi_2\rangle= U(N_t) |\psi_1\rangle$ at position 3 in the Fig.~\ref{fig_OTOC_global_protocol}(a).
	\item Finally, the necessary gates are applied to compute the observable $W$ in the computational basis. For example, if $W = X_i$, an application of the Hadamard gate  before applying projective measurements on qubit $i$ in the computational basis would allow us to compute the expectation value of $\langle W(t) \rangle_{k_0,u}$. The number of shots used is denoted as $N_{M}$.
    \item Likewise, the expectation value of the operator $\langle V^\dagger W(t) V \rangle_{k_0,u}$ is computed with an inclusion of the $V$ operator for the same unitary $u$. The operator is inserted after creating the random state $|\psi_1\rangle$ (position 1 in Fig.~\ref{fig_OTOC_global_protocol}(b)). 
	\item The process is repeated with $N_u$ unitaries for each Trotter step. We choose $N_u$ either to be 600 or 900 based on convergence. The unitary matrices are drawn randomly from CUE.
    \item Finally, an ensemble average of the quantity $\overline{\langle W(t)\rangle_{u, \mathbf{k}_0} \langle V W(t) V\rangle_{u, \mathbf{k}_0}}$ is computed which is a measure of the OTOC.
\end{itemize}

We perform an additional normalization to compute the OTOC with the global protocol
\begin{equation}
O(t)=\frac{1}{ \overline{\langle W(t)\rangle_{u, \mathbf{k}_0}^2}} \overline{\langle W(t)\rangle_{u, \mathbf{k}_0}\langle V W(t) V\rangle_{u, \mathbf{k}_0}}. \label{otoc_def_appendix}
\end{equation}

Following the steps outlined above and using the measurements on circuits like in Fig~\ref{fig_OTOC_global_protocol}, we can look at the correlation of the computed expectation values. This demonstrates the operator spreading due to information scrambling in the system. The speed of the information spreading can thus be measured with the computation of the OTOC using \eqref{otoc_def_appendix}.

\begin{figure*}[!htb]
	\subfloat[\label{corr_t0}]{%
		\includegraphics[width=.33\textwidth]{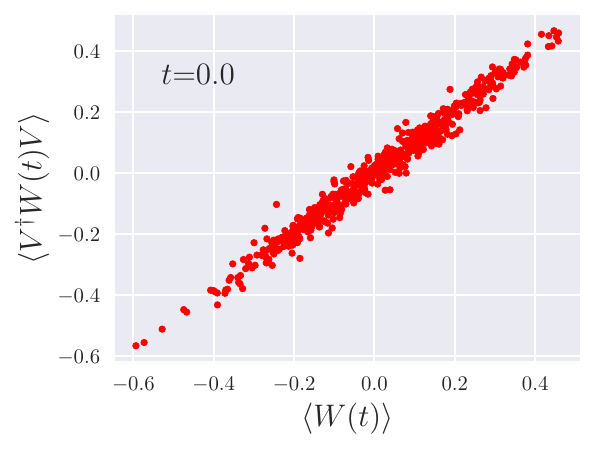}%
	}\hfill
	\subfloat[\label{corr_t1.5}]{%
		\includegraphics[width=.33\textwidth]{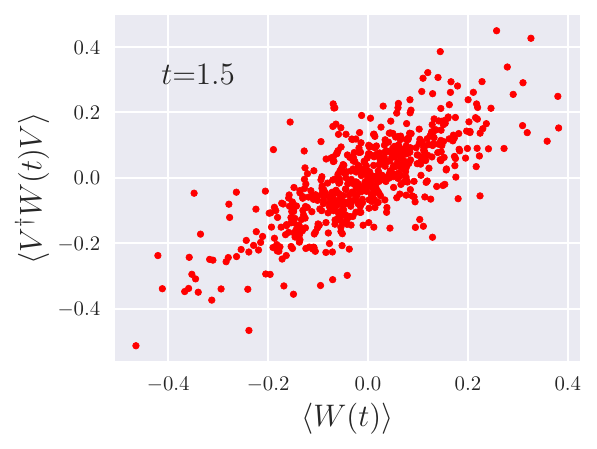}%
	}\hfill
	\subfloat[\label{corr_t3.5}]{%
		\includegraphics[width=.33\textwidth]{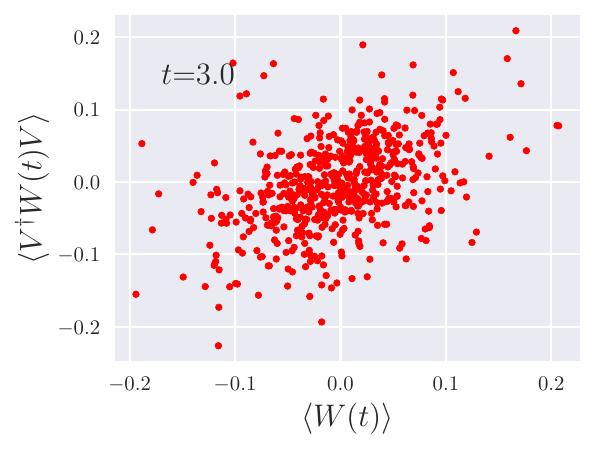}%
	}
	\caption{The change in the correlation of the operators over time in one realization of the $N=6$ SYK model, with $\langle W(t) \rangle=\langle Z_1(t)\rangle$ and $\langle VW(t)V \rangle=\langle Z_0 Z_1(t) Z_0 \rangle$. $Z_i$ denotes a local Pauli-$Z$ operator on qubit $i$.}
	\label{correlations}
\end{figure*}

The protocol introduces computational overhead due to running circuits for many different unitaries $N_u$ for the computation of each data point. However, some salient features of this protocol make this a suitable choice. The protocol is robust against depolarizing error and readout errors since this scales the estimated quantities $\langle W(t)\rangle$ and $\langle V W(t) V\rangle $ at the same rate. This resembles the Twirled readout error extinction (T-Rex) error mitigation technique \cite{van2022model} which involves twirling the circuit with Pauli matrices at the end before measurement. The protocol is also robust against imperfect creation of the random unitaries \cite{Vermersch:2018sru}. It remains to be seen if the protocol can be made more cost-effective by designing the unitaries with approximate construction like that of unitary $t$-design \cite{dankert2009exact,Nakata:2021tmt}. The protocol does not require any ancilla qubits and the backward propagation is not required making it viable for the implementation with the NISQ devices. An extensive study comparing different protocols for the SYK model on hardware is an interesting direction on which we hope to report in the future. 

\section{\label{sec:finiteOTOC}Finite temperature OTOC}

\begin{figure*}[!htb]
	\subfloat[\label{finite1}]{%
		\includegraphics[width=.48\textwidth]{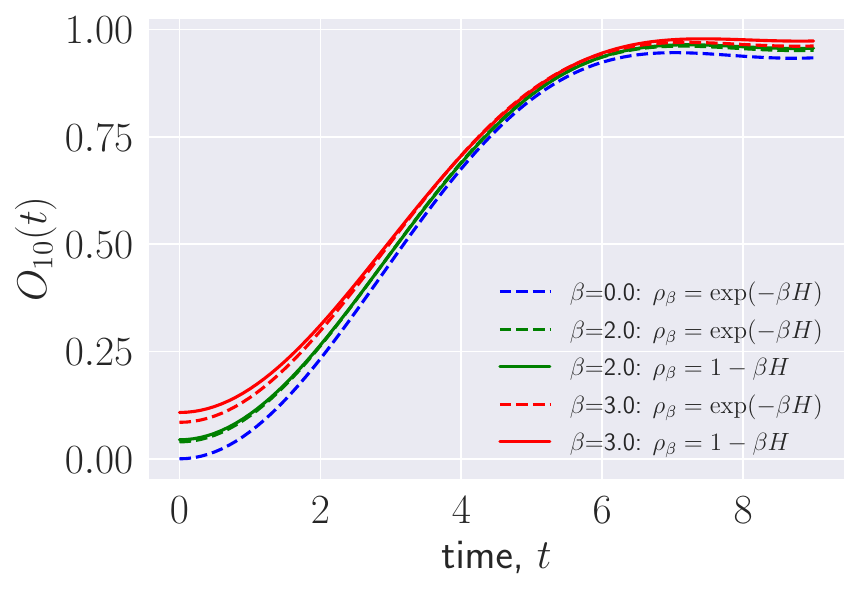}%
	}\hfill
	\subfloat[\label{finite2}]{%
		\includegraphics[width=.48\textwidth]{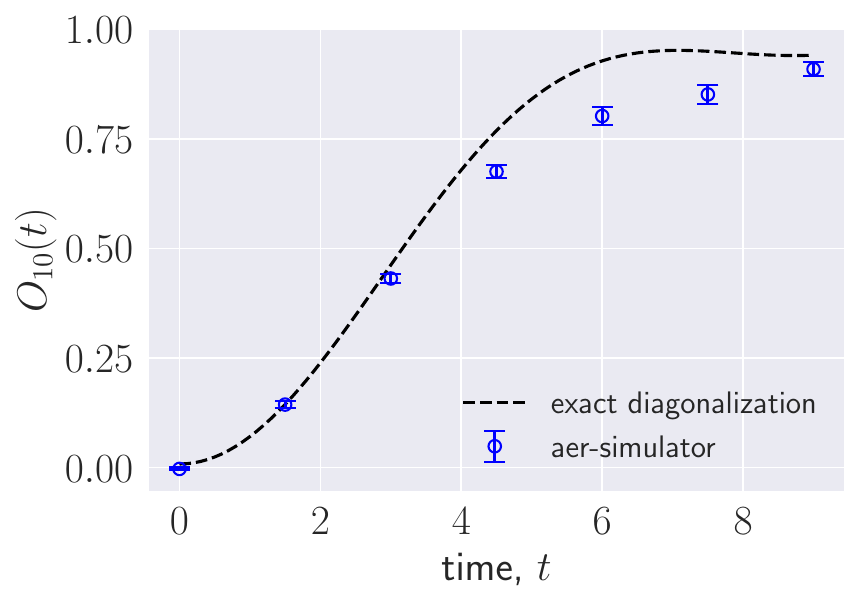}%
	}
	\caption{(a) Comparison of the finite temperature OTOC at with and without the approximation of the thermal density matrix. (b) Finite temperature OTOC using the protocol with first order approximation of the thermal density matrix at $\beta=0.9$ using 2000 global unitaries following Ref.~\cite{Vermersch:2018sru} for $N=6$.}
	\label{correlations}
\end{figure*}

In the main text, we considered out-of-time-ordered correlators at infinite temperatures. In this section, we discuss the extension of the protocol to finite temperature in the high-temperature regime. We consider the symmetrized version of the OTOC defined as:
\begin{equation}
    F\left[\rho_\beta\right](t)=\operatorname{Tr}\left(\rho_\beta^{1 / 4} W(t) \rho_\beta^{1 / 4} V \rho_\beta^{1 / 4} W(t) \rho_\beta^{1 / 4} V\right),
\end{equation}
where $\rho_\beta = e^{-\beta H}/{\mathrm{Tr}(e^{-\beta H})}$ is the normalized thermal density matrix. Employing the high-temperature expansion of the density matrix to first order in $\beta$, we can deduce the following definition of OTOC at non-zero inverse temperature
\begin{equation}
    F\left[\rho_\beta\right](t)= F[\rho_0](t) - \Omega \,\,\overline{\langle W(t)\rangle_{u, k_0}\langle V W(t) V\rangle_{u, k_0}\langle H\rangle_{u, k_0}} + O(\beta^2).
\end{equation}
Here, $F[\rho_0](t)$ is the infinite temperature OTOC which is computed using \eqref{otoc_def_appendix} discussed in the previous section and $\Omega=\frac{\beta}{2}(N_{\mathcal{H}}+1)(N_{\mathcal{H}}+2) $. Since the expectation value of the energy over random unitary ensemble from CUE is time-independent, we are required to compute $\langle H \rangle_{u,k_0}$ only once for a particular inverse temperature.  We observe from Fig.~\ref{finite1} that the correction gives an accurate estimate of OTOC in the regime $0 \leq \beta \lessapprox 2$. However, in this regime, we find this modified definition of OTOC in the finite temperature introduces only a small correction to the infinite temperature ($\beta=0$) OTOC. This small deviation is impossible to capture via simulations of this protocol using current IBM devices. On the other hand, with improvement in the noise characteristics of the device it is evident from Fig.~\ref{finite2} that the protocol can be implemented to calculate finite (high)-temperature OTOCs in small time scales. Our study indirectly, also implies that that a holographic regime or large-$N$ regime can not be reached with this protocol. 


\section{\label{sec:hardware}Additional details about simulations on IBM quantum hardware}

\begin{figure}[H]
    \centering
    \includegraphics[width=0.4\linewidth]{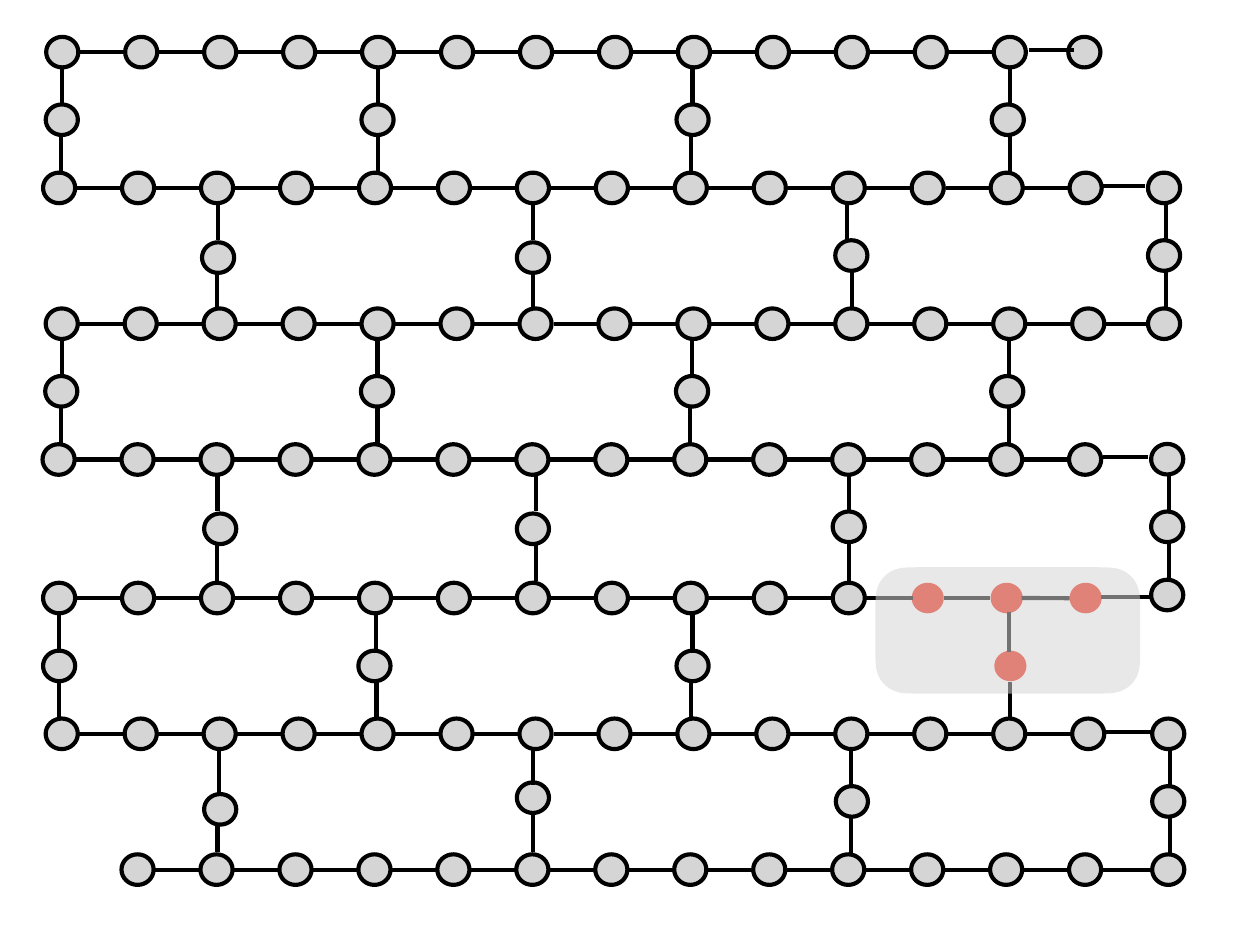}
    \caption{The `heavy-hex' topology of the 127-qubit Eagle processors used in this work. We shade a possible option of four utilized qubits (T-configuration) for $N=8$ which minimizes additional SWAP gates required due to the restricted connectivity of physical qubits.}
    \label{fig:layout_127}
\end{figure}

Our return probability computation was run on \texttt{ibm\_nazca}, \texttt{ibm\_cusco}, and \texttt{ibm\_kyoto}. The OTOC computation was run on \texttt{ibm\_cusco} and \texttt{ibm\_kyoto}. They are all 127 qubit devices with the \texttt{Eagle r3} processor.
The qubit connectivity of these devices is shown
in Fig.~\ref{fig:layout_127}. All these machines use the same set of basis gates which consist of ECR and single-qubit gates $\{\texttt{ID}, \texttt{RZ}, \texttt{SX}, \texttt{X}\}$. 

We first construct the CNOT circuit for one Trotter step and then append the same circuit $N_t-1$ times to give the circuit for $N_t$ Trotter steps. We then transpile the circuits to the above basis gate set with the coupling map of the device we used. Though, we only need $(30)N_{t}$ CNOTs for the entire evolution as discussed in the main text, the actual cost of the two-qubit ECR gate is much higher. This is due to the limited connectivity of the quantum hardware, as shown in Fig.~\ref{fig:layout_127}. 
For $N=6$, we find that about 290 ECR gates (both gate count and circuit depth) are needed for eight Trotter steps, while for $N=8$, we need about 340 ECR gates for two Trotter steps. Thus it poses a challenge to investigate models with non-local interactions with quantum processing units that have restricted topology. The challenges of restricted qubit topology compared to all-to-all qubit connectivity is discussed in the context of the multi-flavor Gross-Neveu model in \cite{Asaduzzaman:2022bpi}. 

Other than minimizing SWAP gates required due to restricted topology, we use the following criteria to map physical qubits to virtual qubits
\begin{enumerate}
    \item Must have low ECR gate error since two-qubit gate errors are the bottleneck in NISQ-era devices.
    \item Have a reasonably high coherence time--- for a larger number of Trotter steps, we typically need $T_{1,2}\sim 200\,\mu$s.
    \item Preferably have a lower ECR gate application time (also sometimes referred to as `gate operation time'). This enables us to perform Trotter evolution to larger times. 
\end{enumerate}

The typical calibration data of some of the devices used in this work are given in Table~\ref{tab:devices_number} 
at the time of use (November/December 2023).
 
\begin{table}[H]
    \centering
    \begin{tabular}{lcccc}
        \toprule
        Device & $\text{EPLG}_{100}$~\cite{McKay:2023nxa} &~~Median $T_1$ (in $\mu$s) &~~Median $T_2$ (in $\mu$s) &~~Median ECR gate error \\
        \hline \hline 
        \texttt{nazca} & $3.2 \times 10^{-2}$ & 186.8 & 115.3& $1.1 \times 10^{-2}$ \\
        \texttt{cusco} & $5.9 \times 10^{-2}$ & 146.6 & 83.3 & $1.74 \times 10^{-2}$ \\
        \texttt{kyoto} & $3.6 \times 10^{-2}$ & 218 & 115.4 & $8.52 \times 10^{-3}$ \\
        \hline 
    \end{tabular}
    \caption{Comparison of some important parameters of the 127-qubit devices we have used in this work.}
    \label{tab:devices_number}
\end{table}
Once good qubits have been chosen, we build the evolution circuits from the Trotter circuits for a given step size. In this work, for the return probability and the OTOC computation, we choose a Trotter step of $dt=1.5$. We can use sufficiently large Trotter step size primarily because of two reasons:
\begin{enumerate}
    \item The relevant quantity to consider for Trotter error estimation is $\left(\overline{J^2_{ijkl}}\right)^{1/2}\,dt\sim\sqrt{3!J^{2}/N^3}\,dt$. Due to the small prefactor for $J=1$, one can choose larger $dt$.
    \item The graph-coloring method reduces the number of terms in the product for Trotterization by grouping the Pauli strings into commuting clusters, thereby reducing the overall error, enabling larger $dt$.
\end{enumerate}
For the return-probability computation, we also build a self-mitigation circuit, where the system is evolved forward by $N_t/2$ steps, then evolved back by the same number of steps. This circuit would have the same structure as the physics circuit. We then Pauli-twirl each ECR gate in the above circuits and generate 75 versions of the physics and self-mitigation circuits. This makes the coherent errors in the circuit appear stochastic, and well-described by the depolarizing channel upon averaging over all the Twirled circuits. The mitigation circuits can be used to estimate the level of depolarizing noise in the circuit. We run these 150 circuits for 2048 shots each. Then, we compute the noise-less value of the observable as discussed in 
the main text.

\raggedright
\bibliography{v3.bib}

\clearpage
\newpage

\end{document}